\documentclass[11pt]{article}
\pdfoutput=1
\newcommand{\be}{\begin{equation}}
\newcommand{\ee}{\end{equation}}
\usepackage{amsmath}
\usepackage{amsfonts}
\usepackage{amssymb}
\usepackage{bm}

\usepackage{cite}
\usepackage{graphicx}
\begin{document}
	
	\begin{center}
		{\Large{\textbf{Mental intervention in quantum scattering of ions without violating conservation laws}}}
		 
		\vspace{2mm}
		 Johann Summhammer\footnote[1]{Email: johann.summhammer@tuwien.ac.at} \\ 
		Technische Universit\"at Wien\\
				Institute of Atomic and Subatomic Physics\\ 
				Stadionallee 2, 1020 Vienna, Austria\\

	\end{center}

\begin{abstract}
There have been several proposals in the past that mind might influence matter by exploiting the randomness of quantum events. Here, calculations are presented how mental selection of quantum mechanical scattering directions of ions in the axon hillock of neuronal cells could influence diffusion and initiate an action potential. Only a few thousand ions would need to be affected. No conservation laws are violated, but a momentary and very small local decrease of temperature should occur, consistent with a quantum mechanically possible but extremely improbable evolution. An estimate of the concurrent violation of the second law of thermodynamics is presented. Some thoughts are given to how this hypothesized mental intervention could be tested.

\end{abstract}

\section{Introduction}

This note is concerned with how mind might influence matter from a quantum mechanical point of view. And in particular, whether such influence violates physical laws. My answer will be NO, with the exception of the second law of thermodynamics, which is not a strict law. The role of mind in quantum mechanics - if it is appreciated at all - is usually considered as that of a \emph{passive} observer. This has been formalized by von Neumann \cite{vonNeumann}: There, the observer becomes aware of the result of an observation, and this new fact replaces the observer's previously calculated probability distribution of possible future results, and consequently the wave function from which it is derived. This sudden change is called the collapse of the wave function. Although the procedure of re-calculating a wave function in the light of new data is a purely logical step, the wave function has often the form of a field in space and time, similar to an electromagnetic field, and this has given rise to decades of debate on how a field can change in an instant. A variety of different interpretations of the wave function have since evolved \cite{Wikipedia-QMinterpretations}. On the minimum view suggested by probability theory and empirical practice, a wave function is a representation of information about a system actually possessed, or assumed to be obtainable, by the one who calculates it. Therefore it is subjective and this has been elaborated in recent years by the development of QBism \cite{Fuchs}, for which a condensed exposition has been given by Mermin \cite{Mermin}. On the other hand, mind's \emph{active} capacity, that of agency or of making a choice, is analyzed less frequently. Notable exceptions are the works of Stapp \cite{Stappfiles}-\cite{Stapp2014} and others, to which I will come shortly. But often this capacity is merely acknowledged in statements like "the observer can choose which measurement to perform". 

While for some philosophical positions the question of mind making a choice may be only a linguistic problem, I do think that mind can make choices which lead to outcomes in future observations, whose probabilities calculated by the usual application of classical and quantum rules, are negligibly small. Choices made by mind need not necessarily be conscious choices. Conscious or not, a choice marks a turn in the further dynamics of a material system. Such a position raises at least two questions. The first is whether a choice will entail a violation of physical laws. The second is, whether it is in principle possible to distinguish in a suspected system the occurrence of a choice from its unperturbed temporal evolution. In particular, whether there should be traces in the human nervous system, or in analogous structures of other creatures or material systems for whom one might postulate the capability of choosing, when a choice is being made. With regard to the first question, already Descartes tried to avoid the violation of conservation laws, and allowed mind only to influence the direction, but not the velocity of movement \cite{Descartes-Morowitz} (classical mechanics was not yet fully developed). Indeed, within classical physics, the deterministic dynamics of material systems leaves no room for a non-material mind to interfere without violating either conservation of momentum, energy or angular momentum. Therefore also a modern approach to the mind-matter problem like that of Augustyn \cite{Augustyn2021}, which does not explicitly expulse determinism, has to allow small violations of physical conservation laws, in order to either inform mind or to let mind inform matter. Wilson, although making use of the quantum mechanical uncertainty relations, argues within the determinism of classical mechanics, and also finds that a non-physical mind would violate conservation laws \cite{Wilson2015}. Below I will look at one of his models from a quantum perspective, and will come to a different conclusion. Now, conservation laws, although originally derived from observations, have become book-keeping laws during the canonization of physics, making it difficult to even formulate a physical problem without assuming them. But their status is being called into question, e.g. by Pitts \cite{Pitts2019, Pitts2021}, whose analysis of mind and conservation laws in classical physics since Leibniz pointed out various weaknesses, for instance, that conservation laws are based on local symmetries, which might sometimes fail. 

However, a detailed description of systems which we might consider as the locus of a mind-matter junction e.g., neuronal cells, will require quantum theory. And quantum theory offers a concept, which is simply not yet there in classical physics, and which might serve as an entry point for mind: Randomness of outcomes as an \emph{intrinsic} feature, and not just as a consequence of insufficient control of experimental parameters. In recent decades, several hypotheses have exploited this feature. Some of these works are reviewed by Smith \cite{Smith2009}. I will mention a few, which are close to what I want to present here. Most prominent is the work of Beck and Eccles, who suggested that "... intention (the volition) becomes neurally effective by momentarily increasing the probability of exocytosis in selected cortical areas..." \cite{BeckEccles1992}. This formulation is a little unfortunate, as changing a probability will generally entail a change of energy. But in the paper it becomes clear that they mean mind selecting an outcome for which the Born rule gives a low probability. I will say more on the difference in the discussion. Chalmers and McQueen are primarily interested in wave function collapse and consciousness, but note that while consciousness in observations is constrained by the Born rule, agentive consciousness might not be, and conscious choices might indeed be more "intelligent" \cite{Chalmers2021}. An unclarity similar to Beck and Eccles remains here, too. Schmidt, in an attempt to understand purported psychokinetic effects, pointed out that quantum physics permits mind making choices without violation of conservation laws, but that there will be a violation of the second law of thermodynamics, in other words a fluctuation or temporary decrease of entropy \cite{Schmidt1987}. As such fluctuations can occur in any finite probabilistic system, Schmidt calls this a weak violation. Morowitz \cite{Morowitz1987} looked at how ions in a neuronal cell would have to be influenced by mind to spark a signal, quite similar to Wilson \cite{Wilson2015}. He also sees a conflict with the second law of thermodynamics, but it would be specific to living systems. For Georgiev consciousness is part of matter and he analyzes "sentience and free will" from an evolutionary angle \cite{Georgiev2024}. He finds that human consciousness is "utterly inexplicable from the principles of classical physics". Regarding choice he remarks "... the act of choosing performs a quantum jump ... In the act of choosing, the quantum probabilities play the role of inherent biases or desires of the quantum agent...". This seems to imply that, contrary to Schmidt and Morowitz, a choice made by mind should not lead to a violation of the second law of thermodynamics. But interestingly, Georgiev cites Dyson, in whose view this appears to be the case \cite{Dyson}: “Our consciousness is not just a passive epiphenomenon carried along by the chemical events in our brains, but is an active agent forcing the molecular complexes to make choices between one quantum state and another. In other words, mind is already inherent in every electron, and the processes of human consciousness differ only in degree but not in kind from the processes of choice between quantum states which we call 'chance' when they are made by electrons”. Stapp has written extensively on how mental intention can result in the desired action \cite{Stappfiles}-\cite{Stapp2014}. He exploits the quantum Zeno effect, which consists in a series of projections of the quantum mechanical wave function of that physical system in the brain, which is relevant for a specific action to occur. By slightly changing the projection operator from one projection to the next, the wave fuction can be kept in an eigenstate and that eigenstate is "bent" step by step until it becomes the one necessary for leading to the desired action. Mind's role thereby is to implement the series of projection operators, and their timing, because each projection is a collapse of the wave function. Stapp calls this role of mind the "holding in place of an action template". His considerations are based on von Neumann's model of measurement in quantum mechanics. Stapp also draws on the work of the psychologist William James, in particular about the willfull effort that upholds the action template. Although there will be no violation of physical laws, the "bending" of an intial wave function into a final one which has negligible overlap with the initial one, means that mind can bring about a physical course which would be very unlikely without mind's intervention. In an example which he gives in \cite{Stapp2008}, an electric oscillation pattern in the brain can be turned by the quantum Zeno effect into one that leads to a desired action. 
An extension of Stapp's ideas to general systems has been presented by Laskey. She introduces the \emph{quantum reducing agent} (QRA) as a system (not necessarily human or even biological), that can cause reductions to some parts of its own physical state, and freely determine both the time and the operator of the reduction \cite{Laskey2018}- \cite{Laskey2019}. This ability is used in combination with the quantum Zeno effect to induce an intended action, just as in Stapp's model. In both Stapp's and Laskey's conceptions the repeated projections done by the quantum Zeno effect lead to an evolution different from an unperturbed system, and for a concrete example it would be interesting to see whether a challenge to the second law of thermodynamics arises here, too.  

\section{How mind could pick the outcomes of scattering events}

Now I will present a quantitative model of how mind might initiate a course of events in a neuronal system without violating energy, momentum or angular momentum conservation. It is similar to Wilson's and Morowitz's, but it focusses on the intrinsic randomness of quantum events.
 
Whenever an organism with a central nervous system initiates an action, it will be caused by firing of neurons in the motor cortex. These neurons themselves will fire because of impulses coming from other neurons higher up in the cortex. If the decision for the action shall be attributed to mind rather than to physiological processes in the organism, mind must be able to cause the firing of one or several neurons early on in this chain. For the present purpose it is not necessary to know which neurons these are, as only a possible mechanism of how such a firing might come about shall be explained. The basic idea is this: Mind shall be able to select a desired direction of the otherwise random diffusion of ions in the region of the axon hillock of a neuronal cell. This changes the local voltage across the cell membrane, which will initiate an action potential travelling along the axon to the neurons downstream. Importantly, mind's interventions must not violate any physical conservation laws like those for energy, momentum or angular momentum. Figure 1a shows a typical neuron and figure 1b shows the voltage across the membrane of the axon during an action potential.

\begin{figure}[h]
	\centering
	\includegraphics[scale=0.44]{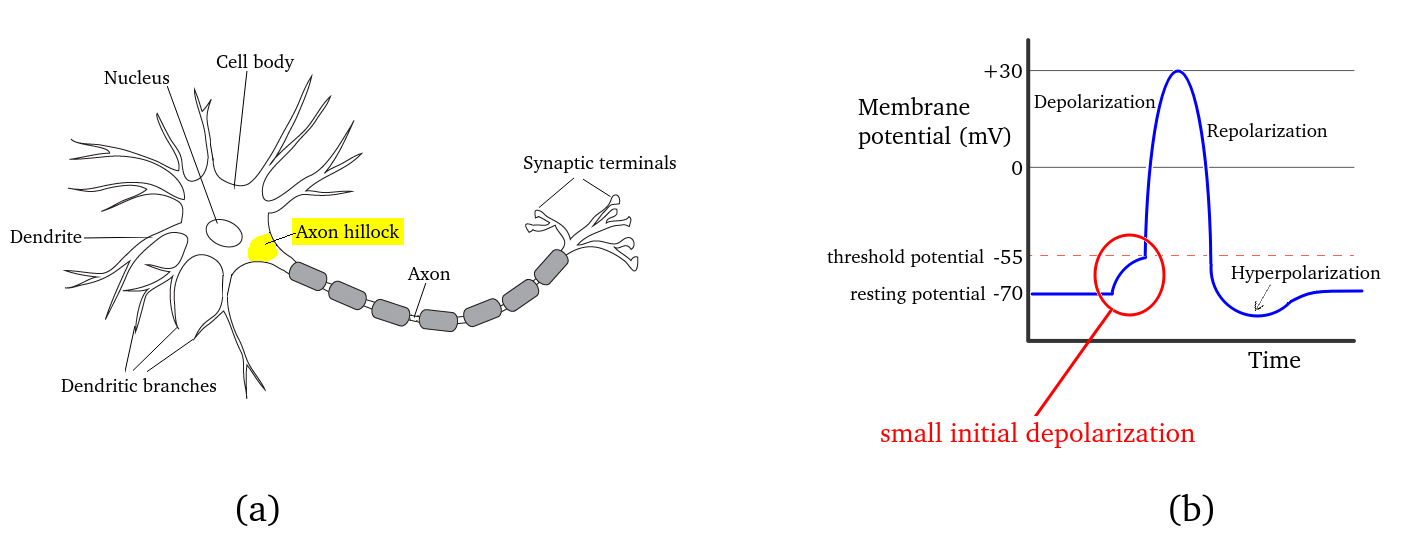}
	\caption{(a) Typical neuron. The axon hillock is where a signal in the form of an action potential begins and propagates along the axon. The present proposal assumes that mind can select scattering directions of ions in this region. (b) Voltage across the membrane anywhere along the axon when an action potential pulse is travelling through. In the resting state the potential inside the axon is around -70 mV relative to the outside. By a depolarization it is raised to about -55 mV which sets off the action potential pulse.}
	\label{}
\end{figure}

\subsection{Details of the model}
Diffusion in the liquid interior of an axon consists of successive scattering processes between ions and the molecules around it, mostly water, or more likely between the hydrated ions and the sourrounding water molecules and water clusters. Any such collision can be broken down into many more interactions between water molecules and an ion, but the relevant aspect is that the whole process must be described quantum mechanically. This means that even if the initial conditions of any two colliding particles are well defined, the outcome of the collision will be a superposition of many possible final states. Each of these possibilities will individually conserve energy, momentum and angular momentum. For simplicity we just look at an ion with initial momentum ${\bm{k}_i}$ and some collison partner with initial momentum ${\bm{K}_i}$, both in the rest frame, such that $\bm{K}_i = -\bm{k}_i$. Then their common final state immediately after the collision will be
\be
\left|\bm{k}_f, \bm{K}_f\right> = \sum_j{w_j \left|\bm{k}_j, \bm{K}_j\right>},
\ee
where in practice the summation may well be an integral over a continuum of possibilities \cite{Renga}. The $w_j$ denote the various probability amplitudes. Momentum conservation ensures $\bm{K}_j = -\bm{k}_j$ for each \emph{j}. This superposition will be extremely short-lived, typically below a picosecond, because the forces of both collision partners to surrounding particles lead to fast decoherence, so that it can be thought of as collapsing almost right away to a reduced superposition or to just a single one of the final possibilities. Without mind's intervention, the result of this reduction is random, but weighted by the probability amplitudes $w_j$. However, if mind intervenes, its desired outcome $j$ is chosen, independent of the value of $w_j$. Figure 2 shows an illustration of an individual scattering event. \footnote{In correct quantum theory decoherence is not the same as collapse of the wave function. Therefore we actually have to think of mind also setting the time of collapse of the wave function right after the interaction of the two particles.}

\begin{figure}[h]
	\centering
	\includegraphics[scale=0.3]{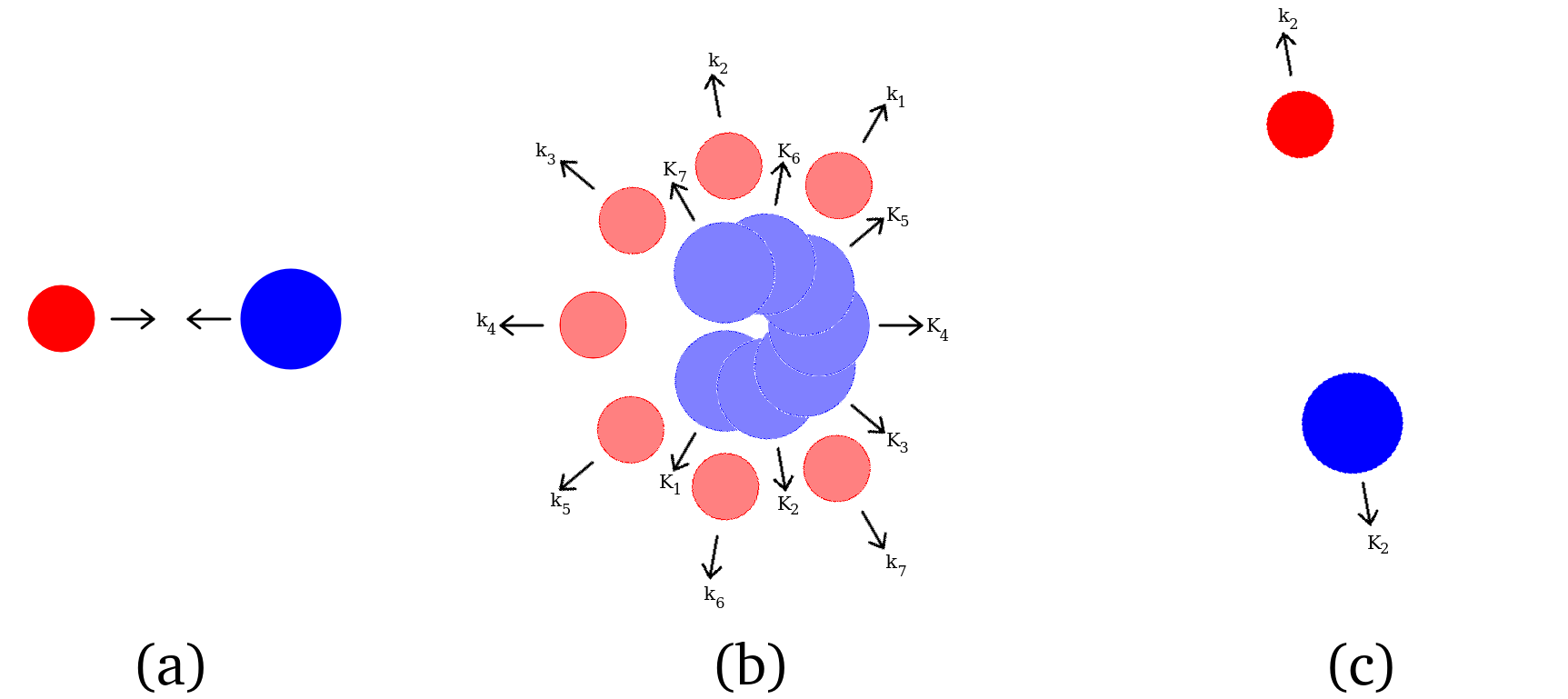}
	\caption{Rest frame view of a scattering event between two particles in the liquid. (a) before the collision, (b) immediately after the collision the joint state of the particles is a superposition of many possibilities, (c) after reduction of the state one of the possibilities becomes factual.}
	\label{}
\end{figure}

Now we shall calculate how many ions must be influenced by mind to diffuse across the cell membrane in the region of the axon hillock in order to elevate the electric potential inside the axon from resting level to the level needed to initiate an action potential pulse. Specifically, we analyze a cylindrical section with a radius $R_a$ of 2 µm and a length of 4 µm as shown in Fig.3a. On both sides of the cell membrane there are many dissolved negative and positive ions (Na$^+$, K$^+$, Cl$^+$, etc.) as well as H$_3$O$^+$, OH$^-$ and charged macromolecules. The concentrations at resting potential of the most important ions for firing of a neuron, Na$^+$ and K$^+$, are listed in Table 1. An estimate of the electric potential difference between the inside and the outside of the membrane section can be obtained by taking the Coulomb potential of each ion and summing over all the ions in a sufficiently large volume, so that extending the summation even further will have negligible influence on the voltage difference. At a point $\bm{r}$ the Coulomb potential is
\be
V(\bm{r})= \frac{1}{4 \pi \epsilon_0 \epsilon} \sum_j \frac{q_j}{\left| \bm{r} - \bm{r}_j \right|}.
\ee
Here, $\epsilon_0$ is the vacuum dielectric constant, $\epsilon$ is the relative dielectric constant in the liquid, and $q_j$ and $\bm{r}_j$ are charge and position of ion $j$, respectively. In order to get a numerical result the cut off for the region with charges outside of the axon has been set to a radius of $R_o = \sqrt{2}R_a$, such that the volume between the membrane and the cut off is the same as that inside the membrane (Fig.3b). The summation in eq.(2) will therefore have to include all ions within a radius of $R_o$ ($\approx$2.83 µm) and the length of the axon section (4 µm). 

\begin{figure}[h!]
	\centering
	\includegraphics[scale=0.50]{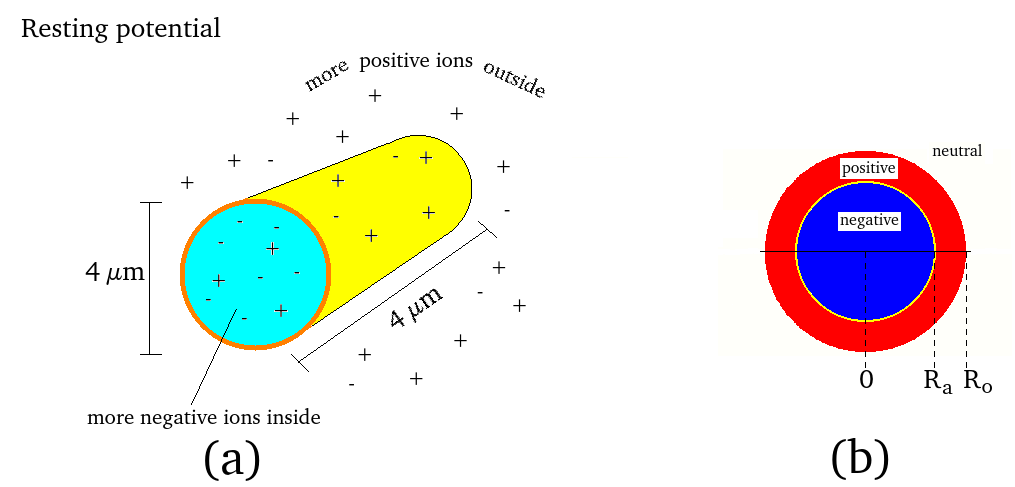}
	\caption{(a) Considered cylindrical section of the axon hillock and distribution of charges under conditions of resting potential. (b) Model of charge distribution. }
	\label{}
\end{figure}

\noindent From Table 1 we see that this would include a very large number of ions. But one has to consider that everywhere inside and outside of the membrane the concentrations of ions of one polarity are almost completely offset by ions of the opposite polarity, otherwise the small physiological potential differences of below 100 mV could not be explained. We can therefore think of the charge distribution at resting potential as consisting only of a small number of singly charged negative ions within the radius $R_a$ and the same number of singly charged positive ions between $R_a$ and $R_o$. We shall also assume that these charges are distributed homogenously in their respective domains. The value of $\epsilon$ in neuronal matter lies in the range of 30-60. Here it was set to $\epsilon$=50. 
\begin{table}[h]
  \begin{center}
    \caption{Concentrations of K$^+$ and Na$^+$ ions inside and outside of a typical neuronal axon at resting potential, diffusion constants D, and mean diffusion distance $\Delta s$ .}
    \label{tab:table1}
    \begin{tabular}{c|c|c|c|c|c|c} 
      & \bf{inside} & \bf{outside} & \bf{inside} & \bf{outside} & \bf{D} & \bf{$\Delta s / 100 ms$} \\ 
      & mmol/l & mmol/l & 10$^6$ ions/µm$^3$ & 10$^6$ ions/µm$^3$ & m$^2$/s & µm\\
      \hline
      K$^+$ & 140 & 5 & 84.3 & 3.0 & 1.96$\times$10$^{-9}$ & 3.7\\
      Na$^+$ & 14 & 145 & 8.4 & 87.3 & 1.33$\times$10$^{-9}$ & 2.5\\ 
    \end{tabular}
  \end{center}
\end{table}

\noindent With this we can calculate how much charge there must be on either side of the membrane to obtain a desired potential difference.
Concretely, an informed guess was made for the number N of negative charges to be placed randomly within the radius $R_a$ and the same number of positive charges were then placed between $R_a$ and $R_o$. Then eq.(2) was used to calculate the potential at many points $\bm{r}$ along a radial line through the center of the cylinder. As the resulting curve $V(\bm{r})$ varied slightly for each calculation due to the random positions of the charges, the calculation was repeated many times and an averaged curve was obtained. The difference between the minimum (inside $R_a$) and the maximum (outside $R_a$) of that curve corresponded to the potential difference of interest. In order to get the desired value, a series of trials with different N had to be made. 
Figure 4 shows the results. For the resting potential (Fig.4a) the total charge within $R_a$ was around 8300 negative ions and as many positive ions between $R_a$ and $R_o$. For the curve of the threshold potential (Fig.4b) it turned out that around 6500 negative ions must be inside $R_a$ and the same number of positive ions between $R_a$ and $R_o$. This implies that raising the electric potential of the 4 µm long cylindrical axon section from resting level to threshold level requires only about 1800 singly charged positive ions like Na$^+$ to diffuse through the membrane from outside to inside. Formally, it could also be achieved if the same number of negative ions diffused from inside to outside. 

\begin{figure}[h!]
	\centering
	\includegraphics[scale=0.72]{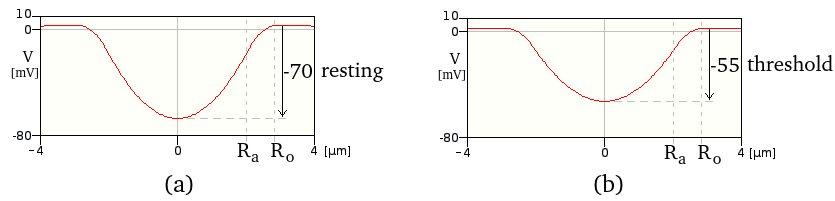}
	\caption{Electric potential $V(\bf{r})$, with $\bf{r}$ along radial line through midpoint of axon section. (a) For charge distribution of resting potential. (b) For charge distribution of threshold potential. }
	\label{}
\end{figure}

\noindent Since the diffusion through the cell membrane at conditions of resting potential implies a diffusion against the balance upheld by the various ion pumps \cite{Stys1995} it is worth looking at possibilities of how the influence of mind could change the electric potential, without ions having to cross the cell membrane. For instance, ions could be made to diffuse \emph{within} the cytoplasm from the neuronal cell body to the axon hillock, or ions could be made to diffuse laterally only \emph{outside} the cell membrane, or a combination thereof. This has been simulated 
in a model shown in Fig.5a. Two axon sections of the same size as shown in Fig.3a were connected, with the left part representing the cell body and the right one the axon hillock. First, conditions of resting potential were created throughout by placing a homogeneous random distribution of positive ions ouside the mebrane and doing the same with negative charges inside. Then different possibilities of shifting ions from right to left or vice versa were investigated. E.g., shifting only charges inside the membrane, or shifting charges only outside, or shifting an equal number both inside and outside. In the latter case it turned out that about 1900 positive ions had to be shifted from right to left outside the membrane, and the same number of negative charges in the same direction within the membrane to raise the electric potential in the center of the right section from resting level of -70 mV to the threshold level of -55 mV. Similar numbers were obtained for the other possibilities. Therefore we can conclude that no matter how we wish to raise the electric potential by about 15 mV in an axon section of typically 4 µm in length and diameter, only a few thousand charges have to diffuse by a few µm in a common direction to initiate an action potential pulse. This is quite a small number which merits further remarks in the discussion.

A critical factor in our hypothesis of mind making factual the desired outcomes of quantum mechanical scattering events is whether the induced change of diffusion direction can initiate a neuronal pulse within the known physiological time intervals. In principle this should be no problem, because virtually all neuronal signalling relies on diffusive processes of ions and small molecules. For such processes the last column of Table 1 gives mean distances which K$^+$ and Na$^+$ diffuse within 100 ms under normal physiological conditions. Note that these few µm are mean distances a random walking ion moves away from its starting point within 100 ms. The actual walk lengths are much greater. Therefore, if mind picks only approximately the direction in which such ions shall move, the distances covered within 100 ms might be many times the normal diffusion distance, and the initiation of a neuronal action potential by mind might actually require less time than the normal physiological process.

\begin{figure}[h!]
	\centering
	\includegraphics[scale=0.72]{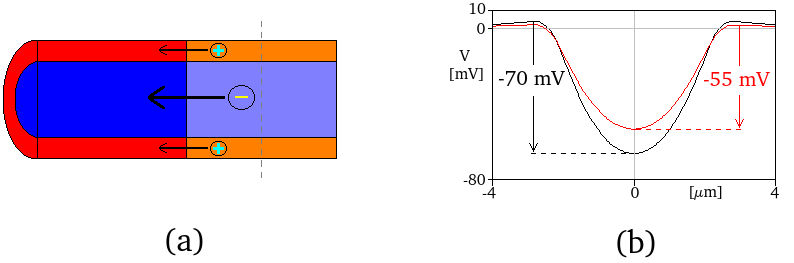}
	\caption{(a) Two sections, in which charges can be shifted laterally from one to the other without crossing the membrane. (b) Electric potential at points along a radial line through the middle of the right section, shown as dashed line in (a); Black: For charge distribution of resting potential. Red: When charges have been moved from left to right to obtain threshold potential.  }
	\label{}
\end{figure}

\subsection{Conservation of energy}
As mentioned above energy is conserved in each scattering event, independent of which of the quantum mechanically possible outcomes becomes factual. Therefore the whole process of billions of successive scattering events also conserves energy. Nevertheless, if mind selects outcomes such that ions are preferentially scattered in a certain direction, a change in the balance between kinetic and potential energy of the ions might build up. This can be seen most easily when looking at our second example in Fig.5. While the first few ions diffusing from right to left will be able to move almost on equipotential lines, after some time a lateral concentration gradient and with it a lateral potential gradient will have developed both inside and outside the membrane so that further ions will have to diffuse against an electric force. This increases their potential energy at the expense of their kinetic energy. The permanent interaction with surrounding molecules leads to an equilibration within the whole volume, resulting in a  decrease of temperature. For the whole volume of around 201µm$^3$ sketched in Fig.5a the total potential energy of the ions at threshold condition (red curve in Fig.5b) is about 2.16x10$^{-18}$ J higher than at resting condition (black curve in Fig.5b). If we assume the specific heat of the intra- and extra-cellular fluid to be more or less the same as that of water (c$_{H2O}$=4184 J/(kg.K)), the temperature of that volume should decrease by about 2.6x10$^{-9}$ °C. Considering naturally occurring temperature gradients within cells \cite{HowardGatenby2019}, this would hardly be measurable. Nevertheless it is interesting to compare this estimate to Wilson's, who conjectured that mind might violate energy conservation if it happens in such a short time interval as to remain below the Heisenberg uncertainty relation \cite{Wilson2015}. He viewed an axon section as an electric capacitor and calculated the difference in stored energy when changing the voltage from resting to threshold level. Taking Wilson's number for the membrane capacity of 1µF/cm$^2$, raising the electric potential in the right section of Fig.5a by 15mV would require 1.13x10$^{-18}$J. This is very close to our estimate, although it should be emphasized that in our model total energy is strictly conserved. 

\section{Discussion}
The analysis presented above touched on several points which are worthy of further comments.

\subsection{Are a few thousand ions really enough?} 
In the calculations we obtained a surprisingly small number of only a few thousand ions to be influenced by mind in a volume on the order of 100 µm$^3$ (or a few tens of ions per µm$^3$) in order to shift the voltage across the neuronal membrane from resting level to threshold level. To put these numbers in perspective, it is perhaps useful to look at how many doping charges need to be added or removed in the two adjacent layers of a silicon diode to obtain different contact voltages. The contact voltage is given by \cite{diode}
$V_c = V_T \ln \left( \frac{N_D N_A}{n_i^2} \right)$, where $V_T$=26mV is the thermal voltage at room temperature, $n_i$=1.45$\times$10$^{10}$ cm$^{-3}$ is the intrinsic carrier concentration for silicon at room temperature \cite{intrinsicSi}, and $N_D$ and $N_A$ are the concentrations of the doping atoms, e.g., phosphorous and boron, respectively. If we assume equal concentrations of phosphorous and boron, and wish to obtain a contact voltage of 500 mV, we would need $N_D=N_A \approx$ 217 µm$^{-3}$. Raising the contact voltage by 15 mV would require to increase the concentration of doping atoms, and thus charges, in each layer by about 72 µm$^{-3}$. This compares favorably to the numbers we found above.

\subsection{Other possibilities for mind's influence}
Our model was concerned with the scattering of ions being influenced by mind. However, there are countless other ways of how mind could intervene to initiate a bodily action. All it takes is the thermal movement of parts which interact with other parts quantum mechanically such that the interaction results in a superposition of possibilities, of which mind can pick the suitable ones and let them become manifest. For instance, Beck and Eccles \cite{BeckEccles1992} have taken the movement of particles towards a potential barrier and mind was to pick the possibility of tunnling through rather than of reflection. In our model, instead of ion scattering in the cellular liquid in general, we could have chosen to look at the \emph{ion channels within the membrane}. There, K$^+$ or Na$^+$ ions are mostly in vibratory motion at specific sites within the channels. One can show that the quantum mechanical wave packets of these ions can have a significant spread within typical coherence times, such that the probability of a particle ending up in a neighbouring site can be higher than expected by a classical molecular dynamics description \cite{Summhammer2020}. Mind might well like to intervene right there and repeatedly collapse wave packets into the neighbouring site, thereby achieving a high charge transfer rate which is very improble otherwise. This would lead to a voltage build up and ultimately to an action potential pulse. Gustav Bernroider remarked that it is also conceivable that mind influences the gates of ion channels \cite{GustavPrivate}. These are macromolecular structures which block the entrance of ions into an ion channel, and which are only opened under the correct physiological conditions, such that ions can stream through the channels and initiate a neuronal firing. One might imagine that the incessant collision-like interactions with surrounding water molecules, which lead to the usual Brownian motion, could be exploited by mind letting more water molecules impact from a certain direction. Yet another possibility was pointed out by Martin Korth \cite{MartinKorthPrivate}, who suggested post-synaptic neurotransmitter receptors as the nexus where mind could initiate a neuronal firing. Binding of a neurotransmitter is subject to quantum fluctuations, and mind could exploit these statistics just as with ion scattering events. The amount of energy which would be redistributed against entropy increase might be even less than in our example with ions. But contrary to ions, which may leave the neuronal cell after a short time, so that mind would have to use different ions in a next intervention at the same neuronal cell, a neurotransmitter receptor is a protein structure, and thus has much longer permanence for its functional role in that specific neuron. In this way mind could be connected to a specific material structure more permanently, which might fit better with an evolutionary picture of the formation of a mind-matter connection. This point indicates that it might be a good approach to look first for those places in the brain which are essential for the passive aspect of mind, that is, for becoming aware of sensory input, and from there seek those neuronal structures which are necessary for a reaction. These would be the places where the active aspect of mind could interfere and implement its intention.

\subsection{How much violation of the second law of thermodynamics?}
If mind shall be able to set the outcomes of quantum measurements as it desires - and this is what our model of mind selecting the scattering directions of colliding particles means - this will lead to a statistical distribution of outcomes of the scattering events, which may deviate strongly from the one expected on purely physical grounds. This raises an issue with the second law of thermodynamics, which states that a system, when left to itself, will tend to maximize its entropy. In the statistical formulation entropy is given by \cite{SecondLawStatistical}, $S=k \ln W$, where $W$ is the number of micro states that are possible for the given system's current macro state and $k$ is Boltzmann's constant. The number $W$ is directly proportional to the probability of the macro state for which it is calculated. During its temporal evolution a system will pass from one macro state to another in a statistical manner, and fluctuations of $W$ to larger or smaller values are possible, just as the fraction of heads in a continued series of coin tosses will not always be 50$\%$, but will go up and down a little.  The second law does not enforce an increase of entropy with every time step. It even allows a decrease of entropy as large as is possible within the rules of probability theory. In an isolated system entropy could go from maximum down to minimum. But this is extremely unlikely to happen, and already for systems of only a few particles such a phenomenon would be a surprise if it actually occurred. 
In our example of Fig.5a, the resting potential of -70 mV in the middle of the right section (indicated by the dashed line) required a homogeneous distribution of 7600 positive charges outside the axon, and 7600 negative charges inside the axon, and the same distribution in the left section. Thus there were altogether 30400 particles. The natural fluctuation (= standard deviation $\sigma$) in each section would be $\sigma = \frac{\sqrt{30400}}{2}$=87.2. In order to achieve threshold potential a total of 3800 particles had to diffuse from right to left, which amounts to about 43$\sigma$. The probability that such a distribution of particles would come about by an accidental fluctuation is on the order of 10$^{-404}$ \cite{Wolfram}. Even though this estimate may be wrong by many orders of magnitude because we neglected physical details, the probability will be so small that, when faced with the actual occurrence of the phenomenon, we could rightfully consider it a violation of the second law of thermodynamics. But the principle question remains, how unlikely an event must be before we can call it a "violation". As in all of empirical science we will have to introduce a confidence level, and in the case of testing for a non-physical mental influence on the physical, we may wish to apply a very strict one. It should be at least as strict as the customary 5$\sigma$ criterion of particle physics, where for extraordinary claims more than 8$\sigma$ have been suggested \cite{Lyons2013}. 

\subsection{Probability distribution versus random events}
In the introduction I mentioned Beck and Eccles \cite{BeckEccles1992} who spoke of mind's intervention as a momentary change of probability of tunneling, while they mean a random event where mind selects tunneling rather than reflection of a particle at a barrier. Ideed, a temporary change of probability of a particle to tunnel through a barrier is physically only possible by changing, for that time, the interaction potential between the particle and the barrier. Since electric charges and spins determine most interactions in biology at the atomic level, mind would have to be able to change constants of nature like charge and magnetic moment of electrons, or bring additional energy into play as suggested by Augustyn \cite{Augustyn2021}. This would definitely be a violation of physical conservation laws. But on the other hand, if mind simply exploited statistical possibilities, as I am arguing here, no such problem arises. In quantum mechanical tunneling there is always a certain probability that a particle can get through a barrier much higher than its energy. Even if the chances were only one in a billion trials, mind could pick that successful trial to happen early on and not at a much later trial. This does not involve a change of probability, nor a violation of conservation of energy. It is like having a Uranium-238 atom suspended in a magneto-optic trap and expecting it never to decay during one's own lifetime, because its lifetime is 4.5 billion years. But it could happen within the first hour. In fact in a Uranium-238 sample of only a few grams quite a number of atoms decay within the first hour of observation. Nevertheless, this raises the issue that theoretically extremely rare events, when observed too frequently, should lead to questioning the underlying theory.
We will discuss this within our model of ion scattering in the neuronal fluid. Suppose we have a subject who self-reports when she/he has made a choice, and in the brain of the subject we have identified a neuronal cell at whose axon hillock firings are initiated without the normal physiological pre-conditions, but in close temporal proximity to the subject's self-report. So we think of these firings as due to mental intervention. We also observe that the same neuron starts many pulses with all the normal pre-conditions in place, but without the subject's self report. After sufficiently long observation we can calculate the fraction of firings due to mental intervention. And this fraction will be significantly higher than expected from a detailed physical modelling of that neuron and its surrounding network assuming no mental intervention. Thus expected probability distribution and observed statistical distribution will not fit together. 
Then we have two options: 
\begin{itemize}
\item{We may accept mental intervention and be content with the coincidence of self-report of the subject of having made a choice and the occurrence of an extremely unlikely diffusion current of ions and corresponding decrease of entropy at the axon hillock of a specific neuron. Thereby we also accept a more or less strong violation of the second law of thermodynamcs.}
\item{We may deny the existence of mental intervention and qualify the self-report of the subject as an utterance of a biological machine. Then we must seek another purely physical explanation. For instance, we might develop a model which states that if the specific neuronal cell is connected in such and such manner to other neuronal cells, the scattering probabilities of ions in the region of its axon hillock will acquire an as yet unexplained directional weighting. Such a model would have to yield a probability distribution for firings in accordance with statistical observations, and although the coincidence of self-report of the subject of having made a choice and a peculiar form of ion diffusion in the axon hillock will remain, no surprising deviations from the second law of thermodynamics could be derived \cite{Reason2016}.}
\end{itemize}
We see that in such a situation a denial of mental intervention may necessitate very complex physical models, possibly with ad hoc hypotheses like making interactions of particles at the atomic level dependent on material configurations at the macroscopic level, that may not be compatible with our current understanding of physics. Interestingly, Morowitz has made a remark which seems to be applicable to both views and which is worth musing on \cite{Morowitz1987}: "...It allows us to think of mind as related to the kind of knowledge a system has of its own state without taking a measurement." 

\subsection{Testing for mental interventions}
Our analysis indicates that mental intervention in the processes of a neuronal network will be noticed as the occurrence of events, which would be so unlikely in a purely quantum physical description, that they can be considered practically impossible. This is much like how we judge the presence of intelligence in our everyday world. When we find an assembly of small twigs in the form of a bird's nest in a tree, we will hardly think it has been made by accidental gusts of wind. Similarly, when we see a stone house on a beach, our first thought will not be that waves have washed up chunks of rock in exactly the right arrangement. More difficult is the case of an archeologist who discovers a piece of sharp flint stone. Is it a product of early humans, or is it an accidental shape of broken stone? The first assumption sees in the shape of the flint stone a sufficiently strong deviation from chance and thus a sufficiently strong violation of the second law of thermodynamics, while the second assumption does not. Therefore, when searching for mental influence, we will have to require that an explanation of the phenomenon by normal physics results in a strong violation of the second law of thermodynamics. Now, where should we look and which experiments can we think of?
 
In our calculations we have assumed that the locus of mental intervention is the nervous system, and in particular we have looked at the diffusion of ions in the region of an axon hillock. Other researchers have suggested other functional units, which is equally conceivable. Strictly speaking, there need not be just one region. There could be many separate regions where mind interferes when a choice is being made, and one should expect different regions for choices leading to different motor reactions. But what is it, that ties mind to these regions? For a non-physical mind it should not make a difference, at which scattering processes it makes selections. When I wish to pick up a candy, mind probably interferes in quantum mechanical scattering somewhere in the nervous system to initiate the proper motions of arm and hand. But shouldn't a non-physical mind be equally capable of interfering in the quantum mechanical scattering directions of the air molecules around the candy and let the candy be lifted on an air cushion right to my mouth? Here one must consider that the information about the candy is received and processed in the nervous system, and options for a choice are represented in some physico-chemical form right there. So it is reasonable to expect, also from an evolutionary viewpoint, that mind intervenes in such places, which are in close physical interaction with these representations rather than any other places. Therefore we should look for mental intervention in the nervous system. But even if we identify one or several neurons whose firing can be linked to mind making a certain choice, the calculations in section 2.2 showed that the local temperature decrease will be only on the order of 10$^{-9}$ °C. This would be totaly drowned by thermal noise, especially as the necessary microscopic probes in or on the axon would be at physiological temperature. The difficulty will be very much the same in the model of Beck and Eccles and the other models mentioned in section 3.2. A direct measurement of voltages might be more promising. There would have to be a voltage clamp at the axon where we think mental intervention is happening, but also at other neurons whose output is feeding into that cell \cite{voltageclamp}. From correlation measurements it might be possible to distinguish normal firings from those due to mental intervention. However, due to thousands of neurons that typically feed into a given neuron in the brain, the practical implementation might be challenging, to say the least \cite{Olah2021}.

But perhaps mind does not only affect the axon hillock of the neuron, or corresponding sites in other models, but also some volume around it. Following this idea, which is admittedly very speculative, it might be possible to observe mental choices in the making, by simply placing appropriate sensors very close to the axon hillock. Such sensors will have to be laid out for detecting a violation of the second law of thermodynamics \emph{directly in their own volume}. This could for instance be a small capsule with an ionic solution, quite like the cerebral fluid, and the quantity to be measured is the voltage across it when the distribution of ions deviates from equilibrium. The electric field from that voltage buildup should possibly enhance either the chance of the neuron's firing or the propagation of the action potential, because this would aid mind in reaching the goal of the choice, and so "it might like" to include the sensor in the volume it acts on. On the other hand, a firing of the neuron should not leave a strong voltage signal in the sensor, so that a voltage buildup in it, when due to a violation of the second law of thermodynamics, can be clearly identified. Naturally, many other working principles for such a sensor can be imagined, in particular solid state sensors, in which only electrons are the moving parts in whose scattering mind would have to intervene. They could be made as small as axon hillocks or even smaller.

\section{Conclusion}
A concrete model has been investigated of how non-physical mind could bring about the firing of a neuron to implement a choice, without violating physical conservation laws. According to this model mind can intervene in the quantum mechanical scattering of ions in the region of the axon hillock of a neuron, which results in the depolarization necessary to set off a neuronal pulse. The envisioned mechanism consists in mind doing collapses of the joint wave function of ions, water molecules and other particles immediately after scattering processes, and letting those scattering directions become actual, which propagate ions into the correct regions within the axon hillock. This leads to the build up of voltage. Calculations show that only a few thousand ions need to be involved. Without mind, such a diffusion of particles is so unlikely, as to be considered impossible, although it is quantum mechanically allowed. Therefore, the intervention of mind would be noticed as a violation of the second law of thermodynamics, specifically as a temporary decrease of temperature in the axon hillock of around 10$^{-9}$ °C. This would be drowned by the physiological temperature fluctuations. Direct measurements of voltages on the relevant neurons also seem to be beyond current technology. But with the additional hypothesis that mind could make similar interventions in a certain volume \emph{around} relevant neurons, sensors which can indicate a violation of the second law of thermodynamics within their own active volume might open paths for experimental tests when placed near such neurons.

\section*{Acknowledgment}

\noindent
I would like to thank Kenneth Augustyn for organizing zoom-meetings around the problem of the mind-matter connection, and I am grateful to him and to the participants of these meetings for quite a number of illuminating contributions.

\begin{flushleft}
	
\end{flushleft}

\end{document}